\documentclass[aps,prd,preprint,floatfix,superscriptaddress,nofootinbib,longbibliography,a4paper]{revtex4-1}
\pdfoutput=1
\usepackage{color}
\usepackage{graphicx}
\usepackage{textcomp}
\usepackage{subfigure}
\usepackage{feynmp}
\usepackage{amsmath,amssymb, array}
\usepackage{enumerate}
\usepackage{slashed}
\usepackage[normalem]{ulem}
\usepackage{hhline}
\usepackage{hyperref}
\hypersetup{colorlinks=true,
    citecolor=blue,
	linkcolor=blue,
	filecolor=magenta,      
	urlcolor=blue}
\newcommand{\mathsym}[1]{{}}

\newcommand{\beqa}{\begin{eqnarray}}
\newcommand{\eeqa}{\end{eqnarray}}
\newcommand{\be}{\begin{equation}}
\newcommand{\ee}{\end{equation}}
\newcommand{\ba}{\begin{array}} 
\newcommand{\ea}{\end{array}}

\begin{document} 
\title{Residual flavour (anti)symmetries at the modular self-dual point and constraints on neutrino masses and mixing}
\bigskip
\author{Monal Kashav}
\email{monalkashav@gmail.com}
\affiliation{Theoretical Physics Division, Physical Research Laboratory, Navarangpura, Ahmedabad-380009, India}
\author{Ketan M. Patel}
\email{ketan.hep@gmail.com}
\affiliation{Theoretical Physics Division, Physical Research Laboratory, Navarangpura, Ahmedabad-380009, India}

\
\begin{abstract}
We explore the implications of symmetries that remain unbroken at the self-dual point $\tau=i$ in modular invariant theories. Assuming that (a) the three generations of lepton doublets transform as an irreducible representation of a finite modular group $\Gamma_N$, and (b) the light neutrino masses arise from the Weinberg operator and are in modular form, we demonstrate that this setup yields a unique residual flavor symmetry or antisymmetry for the neutrinos, depending on the modular weight. In the antisymmetric case, one neutrino is always massless, and the other two can be degenerate if the mass matrix is real. These findings are independent of the level $N$. If the charged leptons are arranged to exhibit an appropriate residual symmetry from the same $\Gamma_N$, they determine a column of the leptonic mixing matrix, leading to specific correlations between the mixing angles and the Dirac CP phase. The presence of residual (anti)symmetries enables the application of standard flavor symmetry techniques to derive these predictions, and we scan all possible $\Gamma_N$ satisfying condition (a). Most solutions yield ${\cal O}(1)$ entries in the fixed column, favouring relatively large lepton mixing.
\end{abstract}

\maketitle

\section{Introduction}
\label{sec:1}
The urge to understand large leptonic mixing has led to some novel and conceptually interesting ideas for imposing horizontal restrictions on the flavours. A relatively recent addition to these schemes is a bottom-up approach based on modular invariance imposed on the supersymmetric version of the Standard Model, extended with mechanisms for non-zero neutrino masses \cite{Feruglio:2017spp}. Modular invariance introduces finite modular groups $\Gamma_N$ as discrete flavour symmetries within the framework, with Yukawa couplings arranged in modular forms. The modular symmetries can also be interpreted as non-linear realizations of ordinary flavour symmetries, where the action of the flavour group on the matter fields is non-linear \cite{Feruglio:2019ybq}. When the modulus acquires a vacuum expectation value (VEV), the Yukawa couplings assume specific numerical values, which can provide constraints on the lepton masses and mixing, depending on the exact implementation of modular invariance and the underlying $\Gamma_N$. Several investigations have been conducted using this principle for different levels $N$; see, for example, \cite{Kobayashi:2018vbk,Penedo:2018nmg,Criado:2018thu,Kobayashi:2018scp,deAnda:2018ecu,Okada:2018yrn,Novichkov:2018yse,Novichkov:2018ovf,Novichkov:2018nkm,Ding:2019gof,King:2019vhv,Wang:2021mkw,deMedeirosVarzielas:2021pug,Novichkov:2019sqv,Ding:2019xna,Ding:2019zxk,deMedeirosVarzielas:2020kji,Ishiguro:2020tmo,Yao:2020qyy,Ding:2020msi,Okada:2020brs,Liu:2020akv,Li:2021buv,Feruglio:2021dte,Kashav:2021zir,Novichkov:2021evw} for some of the initial explorations, and \cite{Ding:2023htn} for a recent review.

In general, the VEV of the modulus $\tau$ breaks the underlying finite group $\Gamma_N$ entirely. While such a value of $\tau$ can be quite successful in reproducing the observed lepton masses and mixing patterns in a given construction, the absence of any apparent residual symmetries at the minimum makes it difficult to understand the qualitative and generic aspects related to the lepton mixing pattern and neutrino mass ordering or hierarchy. For example, one might question whether the solution is truly characteristic of the minima and robust against changes in other model parameters not fixed by modular invariance, or if it's an interplay between these parameters and the VEV, where a small change could lead to significantly different results for the lepton sector observables. This question becomes even more relevant for models plagued by a considerable number of free couplings.

An interesting development in this line of reasoning is the remarkable observation that many models based on modular invariance prefer a modulus value close to $\tau=i$ \cite{Feruglio:2022koo}. The latter is one of the exceptional points, called fixed or self-dual points, where a subset of modular transformations remains unbroken. Following this observation, analyses carried out in the vicinity of $\tau=i$ in \cite{Feruglio:2022koo,Feruglio:2023mii} show some universal behaviour in the predictions for neutrino masses and mixing angles, which do not depend on the explicit models or the choice of modular weight and group $\Gamma_N$, aside from the assumption that the lepton doublets are chosen as irreducible representations of the latter. Both the charged lepton and neutrino mass matrices are shown to possess residual symmetries in the vicinity of the self-dual point, utilizing which the observables of the leptonic sector can be expressed in terms of the expansion parameter $|\tau-i|$, providing an estimate of the size of these observables and their scaling behaviour.

Our present investigation follows a similar approach but with some noteworthy exceptions. We consider the VEV of the modulus as an exact self-dual point, $\tau=i$, without any deviation. Secondly, only the neutrino Yukawa couplings are assumed to be in modular form. Assuming that the lepton doublets transform as an irreducible representation under the finite modular group, these two considerations lead to an exact residual flavour symmetry \cite{Lam:2008rs,Lam:2012ga,Lam:2011ag} or antisymmetry \cite{Joshipura:2015zla,Joshipura:2016hvn} of the neutrino mass matrix. The latter, in particular, leads to a simple constraint on the neutrino mass spectrum, independent of the underlying finite modular group $\Gamma_N$. If the charged leptons are also made invariant under a class of residual symmetries offered by the underlying group, the flavour symmetry or antisymmetry also allows one to predict a column of the lepton mixing matrix. This emergent feature of modular invariant models at the self-dual point is similar to what was studied earlier in the case of ordinary discrete symmetries \cite{Fonseca:2014koa,Joshipura:2016quv}. Utilizing this aspect, we derive all possible predictions for a fixed column offered by each of the suitable $\Gamma_N$.

The results we obtained under the aforementioned assumptions indicate that the modular weight significantly constrains the neutrino mass pattern in certain cases, irrespective of the level $N$ of the finite modular group. In contrast, the fixed column predictions for the lepton mixing do not depend on the weight but only on the level $N$. The majority of solutions obtained for the fixed column suggest that the neutrino mixing parameters are typically large. We identify phenomenologically viable cases and also construct explicit models as illustrations. Modular models with residual symmetries at $\tau=i$ have already been constructed in \cite{Novichkov:2018yse,Novichkov:2018ovf,Novichkov:2018nkm,Ding:2019gof,King:2019vhv,Wang:2021mkw,deMedeirosVarzielas:2021pug} for specific  level $N$, and some of our results have already been observed there. However, our analysis indicates that these are not very model-specific observations but follow from a more general structure of remnant symmetries at the fixed point and can be extended across different levels. Although our approach is less ambitious than that of \cite{Feruglio:2022koo,Feruglio:2023mii}, the more precise predictions and correlations between certain observables might make it  more interesting and practically useful from a model-building perspective.

The remainder of the article is structured as follows. The general aspects of modular invariant theories, with a focus on remnant symmetries at the fixed points, are presented in the next section. In section \ref{sec:3}, we specialize this to bilinears and the case of $\tau=i$. The implications when applied to leptons are discussed in section \ref{sec:4}. In section \ref{sec:5}, we derive a set of predictions for the fixed column of the leptonic mixing matrix. Some simple and explicit model examples are given in section \ref{sec:6}, and the study is summarized in section \ref{sec:7}, with a brief discussion on the implications and validity of the results.

\section{Modular invariance and self-dual point symmetries}
\label{sec:2}
In the bottom-up constructions, the modular invariant supersymmetric theories consist of a set of chiral superfields $\phi_i$ which include the moduli and matter superfields. We consider a simple case with one modulus which is identified as dimensionless $\tau = \phi_0/\Lambda$ with ${\rm Im}(\tau) > 0$. Under the transformations of the homogeneous modular group $\Gamma = SL(2,\mathbb{Z})$, the modulus and ordinary chiral supermultiplets transform as \cite{Ferrara:1989bc,Ferrara:1989qb}
\beqa \label{trans}
\tau &\to & \gamma \tau = \frac{a \tau + b}{c \tau + d}\,, \nonumber \\
\phi_i &\to & (c \tau + d)^{-k_i}\, [\rho_\phi(\gamma)]_{ij}\, \phi_j\,, \eeqa
with $i,j \geq 1$. The transformation parameters $a,b,c,d \in \mathbb{Z}$ and $ad-bc=1$. The modular group $\Gamma$ is infinite, discrete and has a series of infinite normal subgroups $\Gamma(N)$, with positive integer $N$, as defined in \cite{Feruglio:2017spp} and known as principle congruence subgroups. The modular group can be generated by two elements:
\be \label{ST}
S=\left(\ba{cc} 0 & 1 \\ -1 & 0 \ea \right)\,~~{\rm and}~~T=\left(\ba{cc} 1 & 1 \\ 0 & 1 \ea \right)\,,\ee
which satisfy $S^4 = (ST)^3 = \mathbb{I}$. The quotient groups $\Gamma/\Gamma(N) \equiv \Gamma_N$ are finite discrete groups. $\rho_\phi(\gamma)$ in eq. (\ref{trans}) is a unitary representation of $\Gamma_N$ which is reducible in general. $k_i$ are called weights, typically assumed to be integers. A subset of $\phi_i$ transforming as an irreducible representation (irrep) under $\Gamma_N$ must have the same weight.

The invariance of the action of the underlying supersymmetric theory under the above transformations demands the invariance of the superpotential $W(\phi_i)$ and the Kähler potential $K(\phi_i,\overline{\phi}_i)$. The latter is easy to arrange \cite{Ferrara:1989bc,Ferrara:1989qb} while the invariance of the superpotential can lead to some non-trivial constraints. To demonstrate this explicitly, it is more convenient to club the set of $\phi_i$'s which transform as an irrep of $\Gamma_N$. For example, $\phi^{(r)} = (\phi_1,\phi_2,...,\phi_m)$ denotes $r^{\rm th}$ multiplet that transforms as $m$-dimensional irrep. The general superpotential can be expanded in powers of these multiplets. For example, 
\be \label{W}
W \supset \sum_{i_1,...,i_n}\, X_{i_1...i_n}(\tau)\,\, \phi^{(r_1)}_{i_1}...\,\phi^{(r_n)}_{i_n}\,,\ee
where $i_n = 1,...,{\rm dim}(r_n)$. If $\phi^{(r)}$ transform non-trivially under the action of the modular group, $X(\tau)$ also has to transform appropriately to make the above term invariant. Accounting for the transformations of $\phi^{(r)}$ and demanding the invariance of $W$, one finds
\be \label{X_trans}
(c \tau+d)^{-(k_{r_1}+...+k_{r_n})}\,[\rho_{r_1}(\gamma)]_{i_1 j_1}\,...\,[\rho_{r_n}(\gamma)]_{i_n j_n}\,X_{i_1...i_n}(\gamma \tau) = X_{j_1...j_n}(\tau)\,.\ee
Therefore, the functions $X(\tau)$ are required to be modular forms with specific weight and  transformation under the given finite modular group $\Gamma_N$.

In general, $X(\tau)$ transforms as a reducible representation, and it is often convenient to write it in terms of modular forms that are irreps of $\Gamma_N$. Such a decomposition can be parametrized in general as 
\be \label{X_Y}
X_{i_1...i_n}(\tau) = \sum_r\, g_r\, \left[\Gamma^{(r)}_i \right]_{i_1...i_n}\,Y_{i}^{(r)}(\tau)\,,\ee
where $g_r$ are some arbitrary parameters and $[\Gamma^{(r)}_i]_{i_1...i_n}$ are Clebsch-Gordan (CG) coefficients. The latter are numerical factors determined by the tensor-product decomposition. $Y^{(r)}$ is an irreducible representation of $\Gamma_N$ and its elements transform as  
\be \label{Y_trans_simple}
Y^{(r)}_i(\tau)\, \to\, Y^{(r)}_i(\gamma \tau) = (c \tau + d)^{k_{r_Y}}\,[\rho_{r_Y}(\gamma)]_{ij}\,Y^{(r)}_j(\tau)\,,\ee
under the action of the modular group. Substituting eqs. (\ref{X_Y},\ref{Y_trans_simple}) in eq. (\ref{X_trans}), one finds the following two conditions set by modular invariance:
\beqa \label{Y_trans_simp_cond}
k_{r_1} +...+ k_{r_n} &=& k_{r_Y}\,,\nonumber \\
\,[\rho_{r_1}(\gamma)]_{i_1 j_1}\,...\,[\rho_{r_n}(\gamma)]_{i_n j_n}\,[\rho_{r_Y}(\gamma)]_{kl}\, \left[\Gamma^{(r)}_k \right]_{i_1...i_n} &=& \left[\Gamma^{(r)}_l \right]_{j_1...j_n}\,.\eeqa
Notice that the invariance fixes the weight of the modular forms in terms of the sum of the weights of the other fields appearing in the superpotential, and it is independent of the representation $r$ of $Y^{(r)}$. In practice, eq. (\ref{Y_trans_simple}) is used to determine the modular forms of different weights transforming as different irreps of $\Gamma_N$ \cite{Feruglio:2017spp}. Once such a set of $Y^{(r)}$ is obtained, the invariant superpotential is often directly constructed using them.

The modular symmetry is broken spontaneously by the VEV of $\tau$ and the invariance condition, eq. (\ref{X_trans}), is violated in general. However, there are values of the modulus which remain invariant under some smaller set of $\Gamma$ transformations. For example, for certain $\gamma_F$ and $\tau_F$, if
\be \label{FP}
\gamma_F\, \tau_F = \tau_F\,,\ee
then $\tau_F$ is called self-dual or fixed point. Restricting in the fundamental domain of modulus, one finds $\tau_F = i, \exp(2 \pi i/3)\equiv\omega, i\infty$ for $\gamma_F = S, ST$ and $T$, respectively, as the self-dual points. At these points, the full modular group $\Gamma$ is broken but  some of its subgroups, generated by $\gamma_F$, remain unbroken. The latter are called residual symmetries. It turns out that these residual symmetries are Abelian subgroups of $\Gamma$ and they are $Z_4$, $Z_6$ and $Z_\infty$ at  $\tau_F = i, \omega$ and  $i\infty$, respectively \cite{Ding:2023htn}.

Consequently at $\tau=\tau_F$, the invariance condition, eq. (\ref{X_trans}), reduces to 
\be \label{X_trans_fp}
[\rho_{r_1}(\gamma_F)]_{i_1 j_1}\,...\,[\rho_{r_n}(\gamma_F)]_{i_n j_n}\,X_{i_1...i_n}(\tau_F) =  \eta_F^{k_Y}\, X_{j_1...j_n}(\tau_F)\,,\ee
where $\eta_F=-i,\omega^*,1$ for $\tau_F = i, \omega, i\infty$, respectively. Here, we have used the first of eq. (\ref{Y_trans_simp_cond}) and write $k_{r_Y} \equiv k_Y$ since the weight of the modular forms $Y^{(r)}$ is independent of its representation as noted earlier. At the minimum of the modulus, $X$ are essentially the couplings which quantify the strength of interactions between various matter superfields in the superpotential. Therefore, the restrictions implied by eq. (\ref{X_trans_fp}) on $X$ can have phenomenological implications, as we discuss in the subsequent sections.

The constrained form of $X(\tau_F)$ must also follow from the definition eq. (\ref{X_Y}). Indeed, the modular forms which transform as irreps of $\Gamma_N$ are constrained to 
\be \label{Y_fp}
Y^{(r)}_i(\tau_F) = \eta_F^{k_{Y}}\,[\rho_{r_Y}(\gamma_F)]_{ij}\,Y^{(r)}_j(\tau_F)\,,\ee
at the fixed points. Substituting the above in eq. (\ref{X_Y}), the invariance condition (\ref{X_trans_fp}) can be easily verified with the help of eq. (\ref{Y_trans_simp_cond}). The above equation also implies that $Y^{(r)}$ is an eigenvector of $\rho_{r_Y}(\gamma_F)$ with an eigenvalue $\eta_F^{k_Y}$. This can be utilized to determine the modular forms at the fixed points in a relatively easier way for given level $N$ and weight $k_Y$.

\section{Residual symmetries of effective bilinears at $\tau=i$}
\label{sec:3}
Let us now consider a special case of the above in which the effective term in the superpotential is of bilinear type, i.e. it consists of two ordinary chiral supermultiplets whose fermionic components are to be identified with the standard model fermions. It can be parameterized as
\be \label{W1}
W \supset \sum_{a,a^\prime}\,X_{a a^\prime}(\tau)\,\, \phi^{(r_1)}_{a}\,\phi^{(r_2)}_{a^\prime}\, {\cal O}\,,\ee
where $a=1,...,{\rm dim}(r_1)$, $a^\prime=1,...,{\rm dim}(r_2)$ and $\phi^{(r_1)}$, $\phi^{(r_2)}$ as some irreps of the finite modular group $\Gamma_N$. ${\cal O}$ denotes operator containing the other superfields, and it is assumed to be singlet under $\Gamma_N$. It may or may not have non-vanishing weight, and we consider the latter choice for simplicity. When $\tau$ acquires VEV, the term in eq. (\ref{W1}) can be seen as an effective bilinear term, with modular forms $X_{a a^\prime}(\tau)$ playing the role of the Yukawa couplings.

As it can be obtained from eq. (\ref{X_trans_fp}), the effective Yukawa couplings, at $\tau=\tau_F=i$, must obey
\be \label{bl_inv}
\rho_{r_1}^T(S)\,X\,\rho_{r_2}(S) = (- i)^{k_Y}\, X\,,\ee
where $X = X(\tau=i)$ is a matrix of dimension ${\rm dim}(r_1) \times {\rm dim}(r_2)$ while $\rho_{r_1}(S)$ and $\rho_{r_2}(S)$ are square matrices of ${\rm dim}(r_1)$ and  ${\rm dim}(r_2)$, respectively. $\rho_{r_{1,2}}(S)$ are unitary matrices representing the transformation induced by the generator $S$ of the modular group in the representation of $r_{1,2}$. Eq. (\ref{bl_inv}) can be seen as a generalisation of the ordinary invariance condition $\rho_{r_1}^T(S) X \rho_{r_2}(S) = X$.

Following \cite{Rankin_1963}, we consider modular forms of even and non-negative weights only. In this case, 
\be \label{even_weight}
k_Y = k_{r_1}+k_{r_2} \equiv 2k\,~{\rm with}~k\in \mathbb{Z}^+\,.\ee
If the bilinear is of Majorana type, i.e. $\phi^{(r_1)} = \phi^{(r_2)}$, the modular weight $k_Y$ would anyway take only even values. With this, eq. (\ref{bl_inv}) reduces to 
\be \label{bl_inv_2}
\rho_{r_1}^T(S)\,X\,\rho_{r_2}(S) = (-1)^k\, X\,.\ee
The above leads to ordinary flavour symmetry for even values of $k$ or flavour anti-symmetry \cite{Joshipura:2015zla} for odd $k$.

For the even $k_Y$, it is sufficient to consider inhomogeneous finite modular groups as $\Gamma_N$ \cite{Feruglio:2017spp}.  These groups are isomorphic to $PSL(2,N)$ and they are generated by the elements $S$ and $T$ satisfying
\be \label{Gamma_N_cond}
S^2 = (ST)^3 = T^N = \mathbb{I}\,.\ee
The matrix representations of these generators, therefore, satisfy
\be \label{}
[\rho_{r}(S)]^2 = [\rho_r(ST)]^3 = [\rho_r(T)]^N = \mathbb{I}\,,\ee
for any $r$. The first of this implies
\be \label{det}
{\rm Det}(\rho_{r}(S)) = \pm 1\,.\ee

The inhomogeneous finite modular groups with the first few levels $N=2,3,4,5$ are isomorphic to the well-known permutation groups $S_3$, $A_4$, $S_4$ and $A_5$, respectively \cite{deAdelhartToorop:2011re}. For $N>5$, the relations given in eq. (\ref{Gamma_N_cond}) are not sufficient to close the group and additional relations are needed \cite{deAdelhartToorop:2011re}. Nevertheless, the residual symmetry or antisymmetry of the underlying bilinear is determined by $\rho_r(S)$  which corresponds to a $Z_2$ subgroup of $\Gamma_N$ since $[\rho_{r}(S)]^2 = \mathbb{I}$. This residual $Z_2$ symmetry or anti-symmetry can be utilized to constrain the lepton masses and mixing angles, as we show in the next section.

\section{Implications for the leptons}
\label{sec:4}
We now discuss the implications of the fixed point residual symmetry when it is applied to the leptons. Beginning with the neutrino sector, we assume (i) neutrino mass term is of Majorana type, and it arises through an effective Weinberg operator, and (ii) three generations of the lepton doublet superfields, namely $\hat{L}_a$, transform as 3-dimensional faithful and irreducible representation of the underlying finite modular group $\Gamma_N$. In terms of eq. (\ref{W1}), these assumptions essentially imply that $\phi^{(r_1)}=\phi^{(r_2)} \equiv \hat{L}$ and $X(\tau=i) \equiv X_\nu$ is proportional to $3 \times 3$ Majorana mass matrix $M_\nu$ of the light neutrinos. $\rho_{L}(S) \equiv S_\nu$ is  also a $3\times 3$ unitary matrix with $S_\nu^2 = \mathbb{I}$ and a matrix representation of the element of unbroken $Z_2$ symmetry on $\tau=i$. The general result eq. (\ref{bl_inv_2}) then leads to the following constraint on the neutrino mass matrix:
\be \label{Mnu_inv}
S_\nu^T\,M_\nu\,S_\nu =  (-1)^k\,M_\nu\,. \ee
The even (odd) values of $k$ leads to flavour symmetry (antisymmetry) condition on $M_\nu$.

For the odd values of $k$, the above leads to non-trivial constraint on the neutrino masses. Taking the determinant on both the sides and using the fact that ${\rm Det}(S_\nu) = \pm 1$ (see eq. (\ref{det})), one finds 
\be \label{detMnu}
{\rm Det}(M_\nu) = 0\,.\ee
Moreover, since $S_\nu^T = S_\nu$ for all the relevant $\Gamma_N$ (see the next section), we find from eq. (\ref{Mnu_inv}) that
\be \label{trMnu}
 {\rm Tr}(M_\nu) = 0\,,\ee
for the odd $k$. The condition, eq. (\ref{detMnu}), implies at least one massless neutrino. Moreover, if the neutrino mass matrix is real, then eq. (\ref{trMnu}) also leads to a pair of neutrinos with degenerate masses. Clearly, this pattern is not supported by the neutrino oscillation data. If it happens that a slight deviation from $\tau=i$ can alleviate the degeneracy between the pair, without introducing large mass to the otherwise massless state, then this pattern fits better with the inverted ordering in the neutrino mass matrix. It turns out that $M_\nu$ arising from the Weinberg operator with the lowest odd $k$ is real in the models based to the first few levels, $N=3,4,5$. Therefore, these setups would naturally prefer inverted ordering for the values of $\tau$ reasonably close to $\tau_F=i$. Indeed, this can be seen from the solutions obtained in \cite{Criado:2018thu,Novichkov:2018ovf}.

For even values of $k$, conditions in eqs. (\ref{detMnu}, \ref{trMnu}) do not apply and the eigenvalues of $M_\nu$ are not constrained from the residual flavour symmetry. They depend on the values of modular forms at $\tau=i$ and on the decomposition of $X_\nu$ in terms of irreps $Y_i^{(r)}$ of $\Gamma_N$ as indicated in eq. (\ref{X_Y}). Depending on the number of $Y_i^{(r)}$ at the given weight and their values at $\tau=i$, the neutrino masses would still be constrained. However, these constraints cannot be derived solely from the residual symmetries and can be studied only in the model dependent way.

The self-dual point residual symmetry of $M_\nu$ can also lead to restrictions on the matrix which diagonalize it irrespective of the value of $k$. The invariance condition eq. (\ref{Mnu_inv}) can be used to show that
\be \label{Mnu2_inv}
S_\nu^T\, M_\nu M_\nu^\dagger\, S_\nu^* =  M_\nu M_\nu^\dagger\,. \ee
As $S_\nu$ is a hermitian matrix, it can be diagonalized by a single unitary transformation. Let $V_\nu$ be a unitary matrix such that 
\be \label{Snu_diag}
V_\nu^\dagger\,S_\nu\, V_\nu = s_\nu\,.\ee
Since, $S_\nu^2=\mathbb{I}$ and ${\rm Det}(S_\nu) = \pm 1$, the eigenvalues are restricted to be 
\be \label{ev_snu}
s_\nu = {\rm Det}(S_\nu)\, {\rm Diag.}(1,-1,-1)\,.\ee
Therefore, $S_\nu$ must have a pair of degenerate eigenvalues.  Since $M_\nu M_\nu^\dagger$ commutes with $S_\nu$, the matrix which diagonalizes $M_\nu M_\nu^\dagger$  is related $V_\nu$ in the following way \cite{Joshipura:2016quv}
\be \label{}
U_\nu = V_\nu\,P_\nu\,U_{ij}(\theta,\beta)\,,\ee
where $P_\nu$ is a diagonal matrix with undetermined phases and $U_{ij}(\theta,\beta)$ denotes arbitrary rotation in $i$-$j$ plane corresponding to the degenerate eigenvalues in $s_\nu$.

Note that $V_\nu$  can be determined completely from the given $S_\nu$ and it depends only on the choice of finite modular group $\Gamma_N$.  This in turn determine one column of $U_\nu$ in terms of a column in $V_\nu$ up to an overall phase factor, and it does not get altered by the arbitrary rotation $U_{ij}(\theta,\beta)$. Such a column of $V_\nu$ is nothing but the eigenvector of, $S_\nu$ corresponding to the unique eigenvalue. Explicitly, if $c_\nu$ denotes a fixed column of $U_\nu$ then
\be \label{c_nu}
c_\nu = \tilde{c}_\nu\,e^{i \chi}\,\ee
where $\chi$ is an arbitrary phase and $\tilde{c}_\nu$ is an eigenvector of $S_\nu =\rho_{L}(S)$ with eigenvalue $\pm1$ if ${\rm Det}(S_\nu) = \pm 1$.

This restriction on $c_\nu$ emerging from the residual flavour symmetry or anti-symmetry can be utilized to derive predictions for some of the leptonic mixing parameters. However, the latter also depend on the flavour structure of the charged leptons and, therefore, it is desirable that the $3 \times 3$ charged lepton mass matrix $M_l$ also possess some residual symmetries. Let us consider a hermitian combination $M_l M_l^\dagger$ is such that
\be \label{Ml_inv}
T_l^\dagger\,M_l M_l^\dagger\,T_l = M_l M_l^\dagger\,. \ee
Here, $T_l$ is also a matrix representation of some group element of $\Gamma_N$. Again, since $M_l M_l^\dagger$ and $T_l$ are commuting matrices, both can be diagonalized by a common unitary matrix up to an overall diagonal phase matrix. i.e.
\be \label{Ul}
U_l = V_l\,P_l\,,\ee
and $V_l$ given by
\be \label{Vl}
V_l^\dagger\, T_l\,V_l = t_l\,.\ee
$t_l$ is a diagonal matrix, and we require that all its eigenvalues are distinct such that $V_l$ is completely fixed and the relation (\ref{Ul}) holds. This requires $T_l^m=\mathbb{I}$ with, $m\geq3$ and the corresponding residual symmetry group for the charged leptons is $Z_m$.

The residual symmetry of the charged leptons can also be a product of two commuting $Z_2$ symmetries. In this case, $T_l = \{S_{l_1}, S_{l_2}\}$ with $S_{l_i}^2 = \mathbb{I}$ and $[S_{l_1},S_{l_2}] \neq 0$. However, none of $S_{l_{1,2}}$ can be the same $Z_2$ symmetry corresponding to the residual symmetry of the neutrinos at the fixed point $\tau=i$. This is because $T_l$ and $S_\nu$ are required to be non-commuting, otherwise they lead to vanishing mixing in $U_{\rm PMNS}$ \cite{Joshipura:2016quv}. For this reason, we argue that the charged lepton Yukawa couplings are not in modular form otherwise they are bound to possess the residual symmetry identical to that of the neutrinos at $\tau=i$.

Once the residual symmetry of the charged leptons is fixed, one can determine a column of the $U_{\rm PMNS} = U_l^\dagger U_\nu$ matrix using
\be \label{c0}
c_0 = U_l^\dagger\, c_\nu\,= P_l^\dagger V_l^\dagger \tilde{c}_\nu\, e^{i \chi}\,.\ee
Apparently, the absolute values of the elements of $c_0$,
\be \label{c0_mod}
|c_0| = |V_l^\dagger \tilde{c}_\nu|\,,\ee
do not depend on the undetermined phases.  Note that $|c_0|$ is predicted up to the permutations of its rows. The obtained predictions can then be used to find whether the residual symmetries are in agreement with the observed data. The identification of the viable set of residual symmetries  can simplify the task of model building, as we show in the subsequent sections.

\section{Predictions for fixed column of $U_{\rm PMNS}$}
\label{sec:5}
Based on the strategy outlined in the previous section, we derive numerical predictions for $|c_0|$ from the relevant set of finite modular groups $\Gamma_N$. As mentioned earlier, we consider only the groups which possess three-dimensional irreps and this puts a very powerful constraint on the allowed values of $N$. It is already shown in \cite{deAdelhartToorop:2011re} that only a few $\Gamma_N$ possess this property, and it is sufficient to consider $N=3,4,5,7,8,16$ for this purpose. For $N=6$, the group possesses three-dimensional irreps, however, they are not faithful as they can be written as the product of three-dimensional irrep of $\Gamma_3$ and one-dimensional irreps of $\Gamma_2$. Hence, they cannot offer any new predictions for $|c_0|$ other than the ones already obtained from $\Gamma_3$.

To derive the predictions for the fixed column, we first generate all the elements $g$ of the given $\Gamma_N$  from the matrix representations of their generators $S$ and $T$. As discussed earlier, $S_\nu$ is then identified with $\rho(S)$. For $T_l$, we consider all $\rho(g)$ with $\rho(g)^m=\mathbb{I}$, $m \ge 3$ and all three distinct eigenvalues as reasoned earlier. For each combination of $S_\nu$ and $T_l$, we derive, $|c_0|$ as discussed in the previous section. The results in case of each $N$ are discussed below.

\subsection{$N=3$}
\label{subsec:A4}
The finite modular group $\Gamma_3$ is isomorphic to $A_4$ and its 12 elements can be generated by
\be \label{G3_gen}
\rho(S) = \frac{1}{3} \left(\ba{ccc} -1 & 2 & 2\\ 2 & -1 & 2\\ 2 & 2 & -1 \ea \right)\,,~~~\rho(T) = {\rm Diag.} (1, \omega, \omega^2)\,.\ee
We find eight group elements that qualify for $T_l$ in this case. However, all of these lead to a unique prediction for the fixed column:
\be \label{G3_pred}
|c_0| = \left(
\begin{array}{c}
 \frac{1}{\sqrt{3}} \\
 \frac{1}{\sqrt{3}} \\
 \frac{1}{\sqrt{3}} \\
\end{array}
\right)\equiv |c_{\rm TM}|\,. \ee
Comparing with the current $3\sigma$ ranges of the absolute values of the $U_{\rm PMNS}$ matrix \cite{Esteban:2020cvm}, $|c_{\rm TM}|$ can be viably identified as the second column of $U_{\rm PMNS}$. This is a well-known prediction known as trimaximal (TM) mixing \cite{Grimus:2008tt,Albright:2010ap}. In the context of modular models based on $\Gamma_3$, its emergence at $\tau=i$ has already been realised in \cite{Novichkov:2018yse}.

\begin{figure}[t]
\centering
\subfigure{\includegraphics[width=0.42\textwidth]{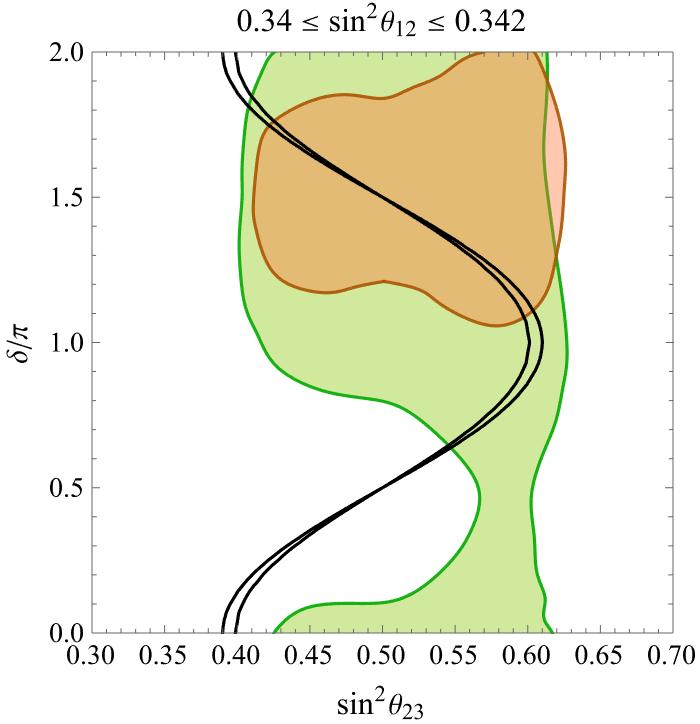}}\hspace*{1.0cm}
\subfigure{\includegraphics[width=0.42\textwidth]{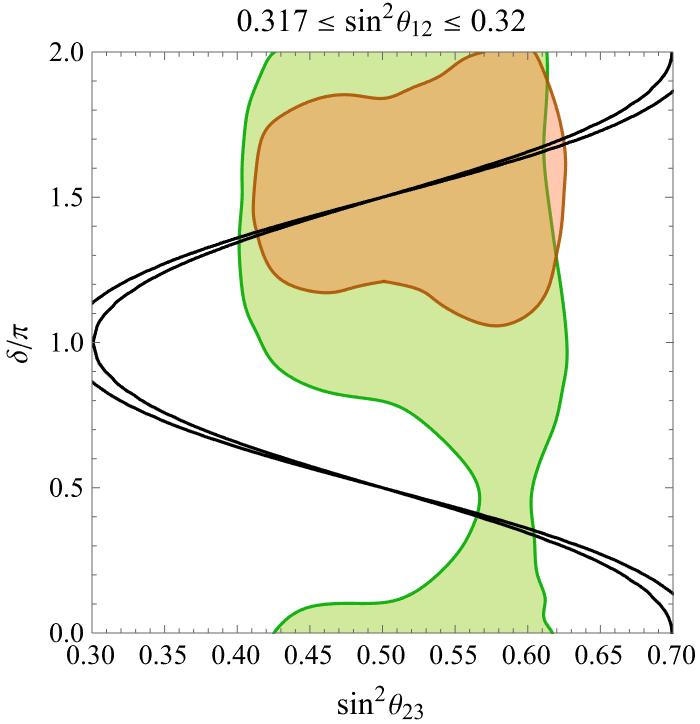}}
\caption{Predictions for a fixed column $|c_{\rm TM}|$ (left panel) and $|c_{\rm TL}|$ (right panel). The predicted range of the solar mixing angle is shown at the top of each panel. The regions enclosed by two black contours represent the theoretically predicted correlation between the atmospheric mixing angle and the Dirac CP phase. The areas shaded in green (red) indicate those allowed by the latest fit of neutrino oscillation data in the most conservative scenario for normal (inverted) ordering.}
\label{fg1}
\end{figure}

Identification of $|c_{\rm TM}|$ as the second column of $U_{\rm PMNS}$ implies the following independent correlations between the mixing angles and Dirac CP phase through the standard parametrization:
\beqa \label{G3_corr} 
\sin^2 \theta_{12} \cos^2 \theta_{13} &=& \frac{1}{3}\,, \nonumber \\
\left|\cos \theta_{12} \cos\theta_{23} - \sin \theta_{12} \sin \theta_{23} \sin \theta_{13}\, e^{i \delta} \right|^2 & = & \frac{1}{3}\,. \eeqa
Consequently, one can find the theoretical predictions for $\theta_{12}$ and $\delta$ provided the values of the other two mixing angles. These predictions are shown in the left panel of Fig. \ref{fg1} allowed by the current $3 \sigma$ ranges of the atmospheric and reactor neutrino mixing angles. 
These specific predictions do not depend on the other parameters and modular weight of the explicit model, as long as the underlying residual symmetries of the charged lepton and neutrino mass matrices are ensured by the vacuum structure of the model. The predictions are applicable to both the orderings in neutrino masses, and the viability of the latter can only be investigated in specific models. An illustration of this is given in the next section.

\subsection{$N=4$}
The group $\Gamma_4 \cong S_4$ has $24$ elements that can be generated using
\be \label{G4_gen}
\rho(S) = \frac{1}{2} \left(\ba{ccc} 0 & \sqrt{2} & \sqrt{2}\\ \sqrt{2} & -1 & 1\\ \sqrt{2} & 1 & -1 \ea \right)\,,~~~\rho(T) = {\rm Diag.} (1, i, -i)\,.\ee
Total $14$ elements are found which can play the role of $T_l$. They lead to three distinct predictions:
\be \label{G4_pred}
\left(
\begin{array}{c}
 \sqrt{\frac{2}{3}} \\
 \frac{1}{\sqrt{6}} \\
 \frac{1}{\sqrt{6}} \\
\end{array}
\right) \equiv |c_{\rm TL}|\,,~~ \left(
\begin{array}{c}
 \frac{1}{\sqrt{2}} \\
 \frac{1}{2} \\
 \frac{1}{2} \\
\end{array}
\right)\,,~~\left(
\begin{array}{c}
 0 \\
 \frac{1}{\sqrt{2}} \\
 \frac{1}{\sqrt{2}} \\
\end{array}
\right)\,.\ee
Only the first prediction is consistent with the neutrino oscillation data if it is identified with the first column of the $U_{\rm PMNS}$ matrix. The pattern is labelled as tri-large (TL) mixing in \cite{Joshipura:2016quv} and TM1 in \cite{Grimus:2008tt}. The other two predictions are also well-known results of $S_4$ group \cite{Joshipura:2016quv} but are not favoured by the data in their exact form. 

The column $|c_{\rm TL}|$ being the first column of the leptonic mixing matrix leads to the correlations:
\beqa \label{G4_corr} 
\cos^2 \theta_{12} \cos^2 \theta_{13} &=& \frac{2}{3}\,, \nonumber \\
\left|\sin \theta_{12} \cos\theta_{23} + \cos \theta_{12} \sin \theta_{23} \sin \theta_{13}\, e^{i \delta} \right|^2 & = & \frac{1}{6}\,. \eeqa
The corresponding predictions for $\sin^2 \theta_{12}$ and $\delta$ are shown in the right panel of Fig. \ref{fg1}. It can be noted that for maximal $\theta_{23}$, both $|c_{\rm TL}|$ and $|c_{\rm TM}|$ predict maximal Dirac CP violation.

\subsection{$N=5$}
The finite modular group $\Gamma_5$  is isomorphic to the permutation group $A_5$. It has $60$ elements that can be generated by
\be \label{G5_gen}
\rho(S) = \frac{1}{\sqrt{5}} \left(\ba{ccc} 1 & \sqrt{2} & \sqrt{2}\\ \sqrt{2} & -\varphi & \frac{1}{\varphi}\\ \sqrt{2} & \frac{1}{\varphi} & -\varphi \ea \right)\,,~~~\rho(T) = {\rm Diag.} (1, \omega_5, \omega_5^4)\,,\ee
with $\varphi = (1+\sqrt{5})/\sqrt{2}$ and $\omega_5 = \exp(2 \pi i/5)$. We find that $44$ elements qualify to be the residual symmetries of the charged leptons in this case. Altogether, they give six distinct predictions:
\be \label{G5_pred1}
\left(
\begin{array}{c}
 0.851 \\
 0.372 \\
 0.372 \\
\end{array}
\right) \equiv |c_{5}^{(1)}|\,,~\left(
\begin{array}{c}
 0.526 \\
 0.602 \\
 0.602 \\
\end{array}
\right)\equiv |c_{5}^{(2)}|\,,~\left(
\begin{array}{c}
 \frac{1}{\sqrt{3}} \\
 \frac{1}{\sqrt{3}} \\
 \frac{1}{\sqrt{3}} \\
\end{array}
\right)\,,\ee
\be \label{G5_pred2}
\left(
\begin{array}{c}
0.934 \\
 0.252 \\
 0.252 \\
\end{array}
\right)\,,~\left(
\begin{array}{c}
 0.357 \\
 0.661 \\
 0.661 \\
\end{array}
\right)\,,~\left(
\begin{array}{c}
 0 \\
 \frac{1}{\sqrt{2}} \\
 \frac{1}{\sqrt{2}} \\
\end{array}
\right)\,.\ee
The solutions displayed in the first line can viably be identified with either first or second column of $U_{\rm PMNS}$. One of them is already obtained from $\Gamma_3$. The solutions shown in the second line above are inconsistent with the data.
\begin{figure}[t]
\centering
\subfigure{\includegraphics[width=0.42\textwidth]{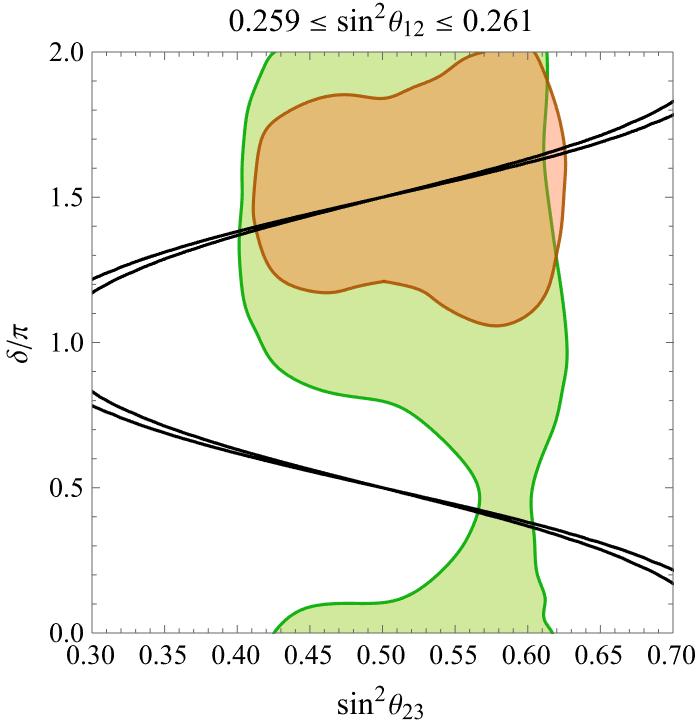}}\hspace*{1.0cm}
\subfigure{\includegraphics[width=0.42\textwidth]{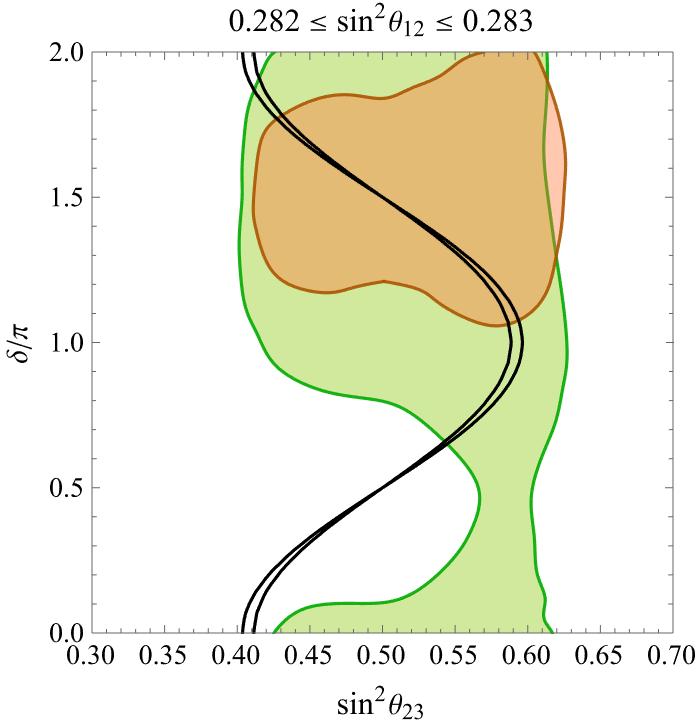}}
\caption{Predictions for a fixed column $|c_{5}^{(1)}|$ (left panel) and $|c_{5}^{(2)}|$ (right panel). For the other details, see caption of Fig. \ref{fg1}.}
\label{fg2}
\end{figure}

We find two new solutions, $|c_{5}^{(1)}|$ and $|c_{5}^{(2)}|$, which are in agreement with the first and second columns of $U_{\rm PMNS}$, respectively. The corresponding predictions are displayed in Fig. \ref{fg2}. Again, they lead to maximal CP violation for $\theta_{23}=\pi/4$. This feature which is seen in case of all solutions (including the ones disfavoured by the data also) so far is due to the fact that the predictions are in agreement with the condition $|U_{\mu i}|=|U_{\tau i}|$ with $i=1$ or $2$. The latter, when followed by all the columns of the lepton mixing matrix $U$, leads to $\theta_{23}= \pi/4$ and $\delta=\pm \pi/2$. $|U_{\mu i}|=|U_{\tau i}|$ for $i=1,2,3$ is the prediction of the generalised version of $\mu$-$\tau$ interchange symmetry \cite{Harrison:2002et}. As shown in \cite{Joshipura:2015dsa}, such a residual symmetry can arise from the discrete subgroups of $O(3)$ and $A_4$, $S_4$ and $A_5$ are the only such groups with three-dimensional irreps. The predicted correlation shown in the right panel of Fig. \ref{fg2} is also seen in the context of a specific model based on modular $A_5$ and with $\tau=i$ in \cite{Novichkov:2018nkm}. As it can be understood now, it follows on the more general ground of residual symmetries offered by $\Gamma_5 \cong A_5$ and  preserved at the self-dual point.

\subsection{$N=7$}
The next relevant finite modular group is $\Gamma_7$ and it is isomorphic to $\Sigma(168)$. The 168 elements of the group can be generated from 
\be \label{G7_gen}
\rho(S) = \frac{2}{\sqrt{7}} \left(\ba{ccc} s_1 & s_2 & s_3\\ s_2 & -s_3 & s_1\\ s_3 & s_1 & -s_2 \ea \right)\,,~~~\rho(T) = {\rm Diag.} (\omega_7^2, \omega_7, \omega_7^4)\,,\ee
where $s_m = \sin(m \pi/7)$ and $\omega_7 = \exp(2 \pi i/7)$. Out of these, $146$ elements are found as possible candidates for $T_l$. They lead to only 8 distinct predictions for the $|c_0|$ given by 
\be \label{G7_pred1}
\left(
\begin{array}{c}
 0.815 \\
 0.363 \\
 0.452 \\
\end{array}
\right) \equiv |c_7^{(1)}|\,,~\left(
\begin{array}{c}
 \sqrt{\frac{2}{3}} \\
 \frac{1}{\sqrt{6}} \\
 \frac{1}{\sqrt{6}} \\
\end{array}
\right)\,,~\left(
\begin{array}{c}
 \frac{1}{\sqrt{3}} \\
 \frac{1}{\sqrt{3}} \\
 \frac{1}{\sqrt{3}} \\
\end{array}
\right)\,, \ee
\be \label{G7_pred2}
\left(
\begin{array}{c}
 0.894 \\
 0.187 \\
 0.408 \\
\end{array}
\right)\,,~\left(
\begin{array}{c}
 0.840 \\
 0.210 \\
 0.5 \\
\end{array}
\right)\,,~~ \left(
\begin{array}{c}
 \frac{1}{\sqrt{2}} \\
 \frac{1}{2} \\
 \frac{1}{2} \\
\end{array}
\right)\,,~\left(
\begin{array}{c}
 0 \\
 \frac{1}{\sqrt{2}} \\
 \frac{1}{\sqrt{2}} \\
\end{array}
\right)\,,~~ \left(
\begin{array}{c}
 0 \\
 0 \\
 1 \\
\end{array}
\right)\,.\ee
The three solutions listed in eq. (\ref{G7_pred1}) above are in agreement with data, out of which two have already been obtained from the smaller $\Gamma_N$. $\Gamma_7$ is the smallest group which offers both the TL and TM solutions.

\begin{figure}[t]
\centering
\subfigure{\includegraphics[width=0.42\textwidth]{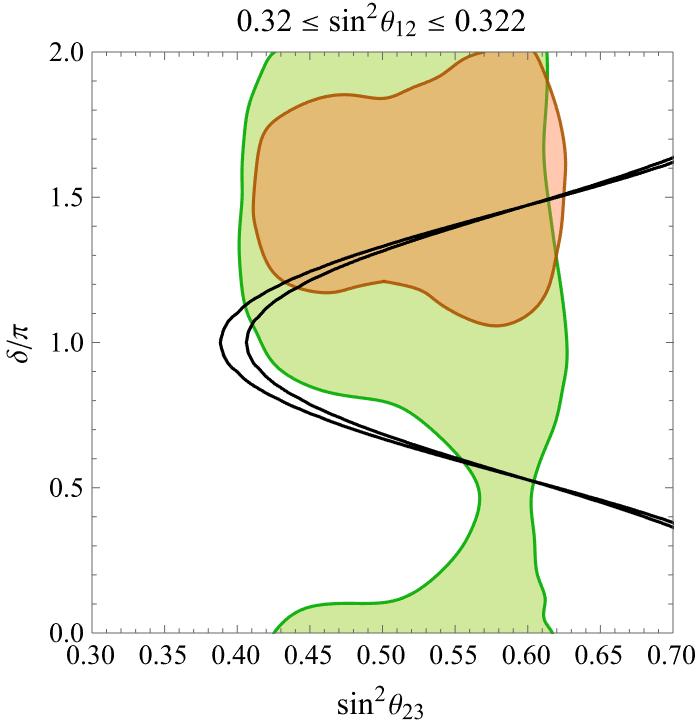}}\hspace*{1.0cm}
\subfigure{\includegraphics[width=0.42\textwidth]{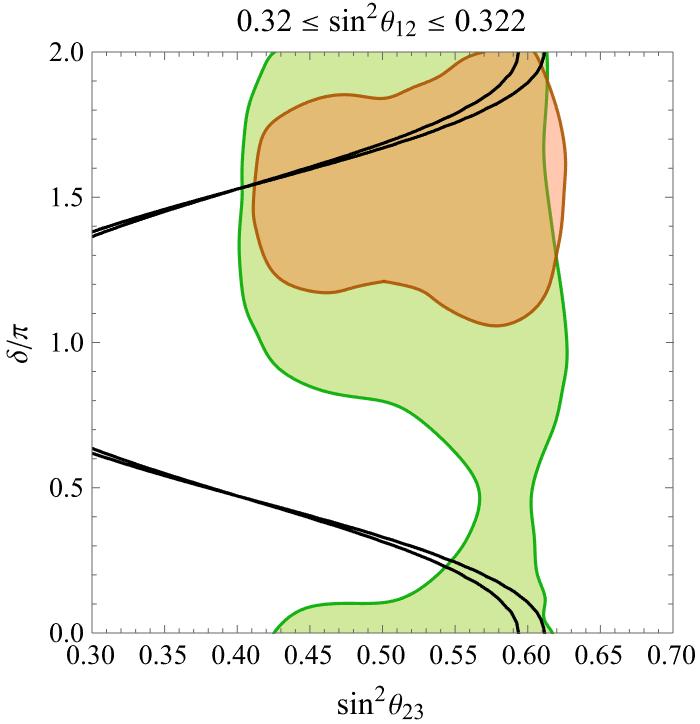}}
\caption{Predictions for a fixed column $|c_{7}^{(1)}|$ (left panel) and its permutation (right panel). For the other details, see caption of Fig. \ref{fg1}.}
\label{fg3}
\end{figure}

The new and viable solutions in this case are $|c_7^{(1)}|$ and its permutation of the second and third rows. Both these can viably be identified with the first column of the leptonic mixing matrix. The corresponding correlations and prediction for the solar mixing angle are displayed in Fig. \ref{fg3}. 

\subsection{$N=8$}
The next group in line is, $\Gamma_8$ which is of order $192$. It contains four three-dimensional irreps, however, they are not faithful. It is sufficient to carry out analysis using its subgroup generated by the three-dimensional representations:
\be \label{G8_gen}
\rho(S) = \frac{1}{2} \left(\ba{ccc} 0 & \sqrt{2} & \sqrt{2}\\ \sqrt{2} & -1 & 1\\ \sqrt{2} & 1 & -1 \ea \right)\,,~~~\rho(T) = {\rm Diag.} (e^{6 \pi i/4}, e^{7 \pi i/4}, e^{3 \pi i/4})\,.\ee
This subgroup has $96$ elements and it is isomorphic to $\Delta(96)$. Therefore, the size of the relevant part of $\Gamma_8$ is actually smaller than $\Gamma_7$. Note that the residual symmetry of the neutrinos at the self-dual point is identical to that in the case of $\Gamma_4$. The residual symmetries of the charge lepton can be different.

This group leads to only $6$ distinct predictions for $|c_0|$ obtained as
\be \label{G8_pred}
\left(
\begin{array}{c}
 0 \\
 0.383 \\
 0.924 \\
\end{array}
\right)\,,~\left(
\begin{array}{c}
 0.211 \\
 0.577 \\
 0.789 \\
\end{array}
\right)\,,~~ \left(
\begin{array}{c}
 \sqrt{\frac{2}{3}} \\
 \frac{1}{\sqrt{6}} \\
 \frac{1}{\sqrt{6}} \\
\end{array}
\right)\,,~~\left(
\begin{array}{c}
 \frac{1}{\sqrt{2}} \\
 \frac{1}{2} \\
 \frac{1}{2} \\
\end{array}
\right)\,,~\left(
\begin{array}{c}
 0 \\
 \frac{1}{\sqrt{2}} \\
 \frac{1}{\sqrt{2}} \\
\end{array}
\right)\,,~~ \left(
\begin{array}{c}
 0 \\
 0 \\
 1 \\
\end{array}
\right)\,.\ee
No new viable solution is found in this case. The group, therefore, do not seem interesting from the perspective of the residual symmetries.

\subsection{$N=16$}
Finally, the last group qualifying our criteria is $\Gamma_{16}$ and just like the previous case, only a subgroup of it is relevant for analysing the fixed column predictions. The latter is generated by
\be \label{G16_gen}
\rho(S) = \frac{1}{2} \left(\ba{ccc} 0 & \sqrt{2} & \sqrt{2}\\ \sqrt{2} & -1 & 1\\ \sqrt{2} & 1 & -1 \ea \right)\,,~~~\rho(T) = {\rm Diag.} (e^{14 \pi i/8}, e^{5 \pi i/8}, e^{13 \pi i/8})\,,\ee
and has $384$ group elements. It is found to be isomorphic to $\Delta(384)$. Despite its big size, it leads to only $10$ distinct predictions for $|c_0|$ out of which only two are in agreement with the data. They are:
\be \label{G16_pred1}
\left(
\begin{array}{c}
 0.810 \\
 0.312 \\
 0.497 \\
\end{array}
\right) \equiv |c_{16}^{(1)}|\,,~\left(
\begin{array}{c}
 \sqrt{\frac{2}{3}} \\
 \frac{1}{\sqrt{6}} \\
 \frac{1}{\sqrt{6}} \\
\end{array}
\right)\,.\ee
For the completeness, we also list the other eight solutions:
\be \label{G16_pred2}
\left(
\begin{array}{c}
 0 \\
 0.195 \\
 0.981 \\
\end{array}
\right)\,,~\left(
\begin{array}{c}
 0.107 \\
 0.648 \\
 0.754 \\
\end{array}
\right)\,,~~ \left(
\begin{array}{c}
 0 \\
 0.383 \\
 0.924 \\
\end{array}
\right)\,,~\left(
\begin{array}{c}
 0.211 \\
 0.577 \\
 0.789 \\
\end{array}
\right)\,,\ee
\be \label{G16_pred3}
\left(
\begin{array}{c}
 0 \\
 0.556 \\
 0.831 \\
\end{array}
\right)\,,~~ \left(
\begin{array}{c}
 \frac{1}{\sqrt{2}} \\
 \frac{1}{2} \\
 \frac{1}{2} \\
\end{array}
\right)\,,~\left(
\begin{array}{c}
 0 \\
 \frac{1}{\sqrt{2}} \\
 \frac{1}{\sqrt{2}} \\
\end{array}
\right)\,,~~ \left(
\begin{array}{c}
 0 \\
 0 \\
 1 \\
\end{array}
\right)\,.\ee
It provides all the solutions offered by $\Gamma_8$ and four new ones.

\begin{figure}[t]
\centering
\subfigure{\includegraphics[width=0.42\textwidth]{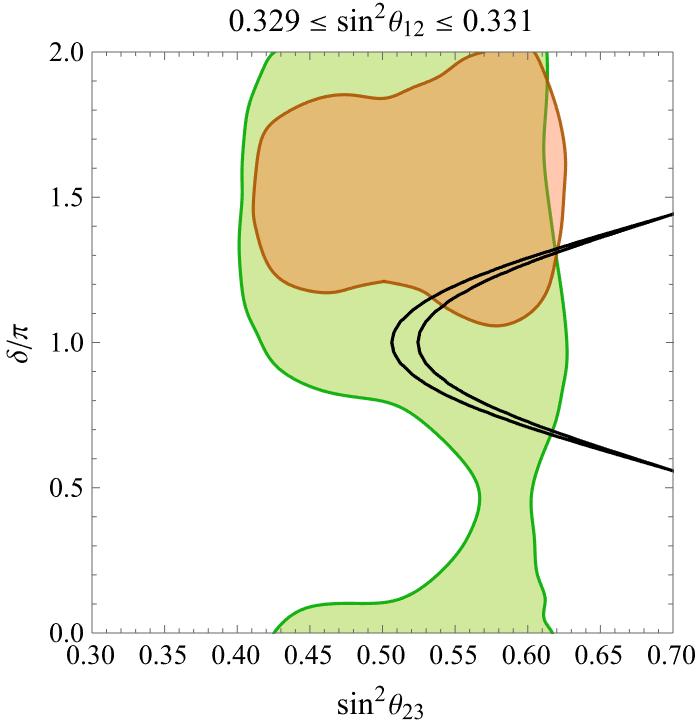}}\hspace*{1.0cm}
\subfigure{\includegraphics[width=0.42\textwidth]{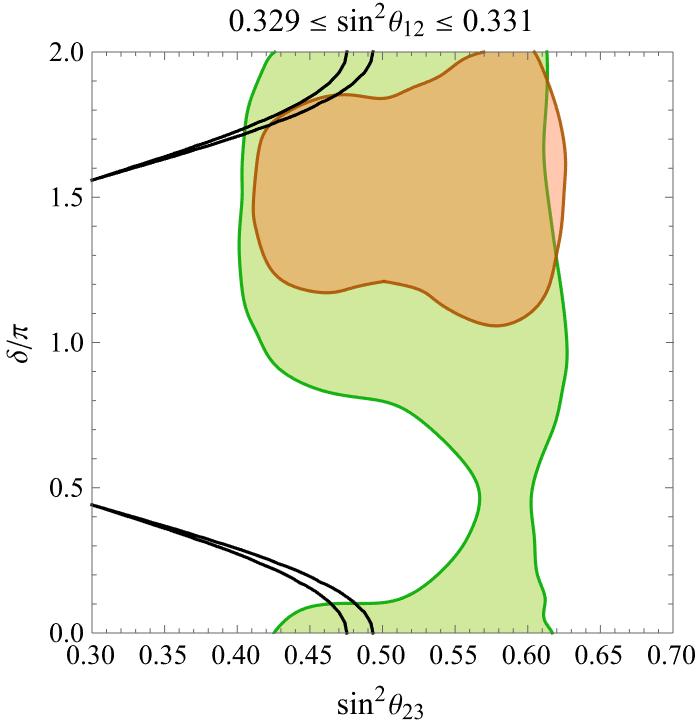}}
\caption{Predictions for a fixed column $|c_{16}^{(1)}|$ (left panel) and its permutation (right panel). For the other details, see caption of Fig. \ref{fg1}.}
\label{fg4}
\end{figure}

As it can be seen, the novel and viable solutions are offered by only $|c_{16}^{(1)}|$ and permutation of its second and third row. They qualify to be the first column of $U_{\rm PMNS}$. The corresponding predictions are given in Fig. \ref{fg4}.
The predictions are quite distinct from the other solutions. The maximal CP violation is excluded entirely in this case.

Altogether, the residual symmetries of the modular neutrino mass matrix at $\tau = i$ and the non-modular charged lepton mass matrix yield five predictions that can be associated with either the first or second column of the $U_{\rm PMNS}$ matrix. This results in five sets of predictions for the solar mixing angles and eight sets of correlations between the atmospheric mixing angle and the Dirac CP phase, as shown in Figs. \ref{fg1}-\ref{fg4}. These findings not only reproduce some known results previously obtained in the context of specific models based on $\Gamma_3$ \cite{Novichkov:2018yse,Ding:2019gof}, $\Gamma_4$ \cite{Novichkov:2018ovf,King:2019vhv}, and $\Gamma_5$ \cite{Novichkov:2018nkm,Wang:2021mkw}, but they also complete the list of all possible solutions emerging from the relevant finite modular groups. As long as the underlying residual symmetries of the neutrino and charged leptons are ensured, these predictions do not depend on the other model parameters or the weights. This is demonstrated in the next section using examples based on some of the simplest models.

\section{Model illustrations}
\label{sec:6}
We present example models based on $N=3, 4$ and for two smallest values of weights to demonstrate explicitly the application of the residual symmetry or antisymmetry.  The models for other values of $N$ with desired residual symmetries can be constructed in the similar way. 

\subsection{Model based on $\Gamma_{3} \cong A_{4}$}
This is the most widely studied finite modular group because of its optimum size. All the essential details about it can be found in \cite{Feruglio:2017spp,Ding:2023htn}. Choosing three generations of lepton doublet superfields $\hat{L}_a$ as $A_4$ triplet, we consider the singlets $\hat{e}^c$, $\hat{\mu}^c$ and $\hat{\tau}^c$ as ${\bf 1}$, ${\bf 1}^{\prime \prime}$ and ${\bf 1}^\prime$, respectively. A flavon $\hat{\Phi}$ is introduced, whose role is to induce $A_4$ breaking in the charged lepton sector in such a way that it leaves its $Z_3$ subgroup unbroken. The $A_4$ and weight assignments are shown in Table \ref{tab1}.  
\begin{table}[h!]
    \centering
    \begin{tabular}{c c c c c c c}
    \hline
    \hline
 ~~~~~~   & ~~~$\hat{L}$~~~ & ~~~$\hat{e}^c$~~~ & ~~~$\hat{\mu}^c$~~~ & ~~~$\hat{\tau}^c$~~~ & ~~~$\hat{H}_{u,d}$~~~ & ~~~$\hat{\Phi}$~~~ \\
     \hline
    $A_4$ & ${\bf 3}$ & ${\bf 1}$ & ${\bf 1}^{\prime \prime}$ & ${\bf 1}^\prime$ & ${\bf 1}$ & ${\bf 3}$ \\
   weight & $k_L$ & $k_{l^c}$ & $k_{l^c}$ & $k_{l^c}$& 0 & $-(k_L+k_{l^c})$  \\
    \hline
    \hline
    \end{tabular}
    \caption{Field content, choice of weights and charge assignments under $\Gamma_{3} \cong A_{4}$.}
    \label{tab1}
\end{table}

The gauge and modular invariant charged lepton superpotential can be written as:
\be \label{}
W_l = \frac{1}{\Lambda} \left(\alpha\,(\hat{L} \hat{\Phi})_{\bf 1}\, \hat{H}_d\, \hat{e}^c + \beta\,(\hat{L} \hat{\Phi})_{\bf 1^{\prime}}\, \hat{H}_d\, \hat{\mu}^c + \gamma\,(\hat{L} \hat{\Phi})_{\bf 1^{\prime \prime}}\, \hat{H}_d\, \hat{\tau}^c\right)\,. \ee
As discussed in section \ref{subsec:A4}, any residual $Z_3$ symmetry for the charged leptons  would lead to desired prediction of TM  mixing as the fixed column. Therefore, we choose $T_l = \rho(T)$ as shown in eq. (\ref{G3_gen}). The flavon VEV configuration invariant under this is obtained as $\langle \Phi \rangle = (v_{\Phi},0,0)^T$. The resultant charged lepton mass matrix is given by
\be \label{Ml_A4}
M_{l} = \frac{v_\Phi v_{d}}{\Lambda}  
\begin{pmatrix}
\alpha & 0  & 0\\
0 & \beta & 0 \\
0 & 0  & \gamma \\
\end{pmatrix}\,. \ee
The desired values of the charged lepton masses can be reproduced using the freedom offered by  $\alpha$, $\beta$ and $\gamma$.

\subsubsection{Weinberg operator with odd $k_L$}
For the lowest weight $k_L = 1$, the Weinberg operator for the neutrino masses can be written using the general parametrization, eq. (\ref{X_Y}), as
\be \label{Wnu_w2_A4}
W_\nu = \frac{1}{\Lambda}\,g_1\,Y^{(2)}_{\bf 3}(\tau)\,(\hat{L} \hat{L})_{\bf 3}\,H_u\,H_u\,.\ee
The coupling $g_1$ can be chosen real without the loss of generality. $Y^{(2)}_{\bf 3}(\tau) = (Y_1(\tau),Y_2(\tau),Y_3(\tau))^T$ is the modular Yukawa multiplet which transforms as $A_4$ triplet, and it is the only multiplet present for weight $k_Y = 2 k_L= 2$. It can be seen from eq. (\ref{Y_fp}) that $Y^{(2)}_{\bf 3}(\tau)|_{\tau=i}$ is an eigenvector of $\rho(S)$ with eigenvalue $-1$.  Since $\rho(S)$ is real and symmetric, $Y^{(2)}_{\bf 3}(\tau)|_{\tau=i}$ is also real. Therefore, the  neutrino mass matrix obtained from eq. (\ref{Wnu_w2_A4}) is real. Following the arguments based on residual antisymmetry in section \ref{sec:3}, it must corresponds to two degenerate and a massless neutrino.

Explicitly, the neutrino mass matrix obtained from eq. (\ref{Wnu_w2_A4}) is given by
\be \label{Mnu_A4_w2}
M_\nu = \frac{g_1 v_u^2}{\Lambda} \left(\ba{ccc} 2Y_1 & -Y_3 & -Y_2 \\ -Y_3 & 2 Y_2 & -Y_1 \\ -Y_2 & - Y_1 & 2 Y_3 \ea \right) \Bigg|_{\tau=i}\,.\ee
At the self-dual point $\tau=i$, the modular functions $Y_{1,2,3}(\tau)$ are given by \cite{Ding:2023htn}
\be \label{Y_i_A4}
Y_2(i) = (1-\sqrt{3})\,Y_1(i)\,,~~~Y_3(i) = (-2 + \sqrt{3})\,Y_1(i)\,. \ee
Substituting the above in $M_\nu$, we indeed find a massless and a pair of degenerate neutrinos.

For higher odd weights ($k_L = 3, 5, \dots$), multiple irreps of Yukawa modular multiplets can exist. Consequently, the neutrino mass matrix may consist of multiple terms and can be complex, even if the modular functions are real at $\tau = i$. As ensured by the residual antisymmetry, one still finds ${\rm Det}(M_\nu) = 0$ and ${\rm Tr}(M_\nu) = 0$, though the latter does not necessarily imply a degenerate pair of neutrinos.

\subsubsection{Weinberg operator with even $k_L$}
Consider $k_L=2$ as the simplest possibility. The Weinberg operator must consist of modular forms of weight $k_Y= 2 k_L = 4$. There are three such non-vanishing irreps: $Y^{(4)}_{\bf 1}$, $Y^{(4)}_{\bf 1^\prime}$ and $Y^{(4)}_{\bf 3}$ \cite{Ding:2023htn}.  Using the expansion eq. (\ref{X_Y}), the neutrino mass matrix can be obtained as
\beqa \label{Mnu_A4_w4}
M_\nu &=& \frac{v_u^2}{\Lambda} g_1\Bigg[\left( \ba{ccc} Y_{\bf 1}^{(4)} & 0 & 0\\
 0 &  0 &Y_{\bf 1}^{(4)}   \\
0 &  Y_{\bf 1}^{(4)} & 0\\ \ea \right) + \frac{g_2}{g_1} \left( \ba{ccc} 0 & 0 & Y_{\bf 1^\prime}^{(4)}\\
 0 &  Y_{\bf 1^\prime}^{(4)} & 0 \\
Y_{\bf 1^\prime}^{(4)} &  0 & 0\\ \ea \right) \Big. \nonumber \\
&+& \Big. \frac{g_3}{g_1}  \left(\ba{ccc} 2(Y_{\bf 3}^{(4)})_1 & -(Y_{\bf 3}^{(4)})_3 & -(Y_{\bf 3}^{(4)})_2 \\ -(Y_{\bf 3}^{(4)})_3 & 2 (Y_{\bf 3}^{(4)})_2 & -(Y_{\bf 3}^{(4)})_1 \\ -(Y_{\bf 3}^{(4)})_2 & - (Y_{\bf 3}^{(4)})_1 & 2 (Y_{\bf 3}^{(4)})_3 \ea \right) \Bigg]\Bigg|_{\tau=i}\,.\eeqa
The modular multiplets  $Y^{(4)}_{\bf 1}$, $Y^{(4)}_{\bf 1^\prime}$ and $Y^{(4)}_{\bf 3}$ can be expressed in terms of the fundamental modular functions $Y_{1,2,3}(\tau)$. The explicit expressions can be found in \cite{Ding:2023htn}. Substituting them in the above at $\tau=i$, one can verify that the $M_\nu$ satisfies the residual symmetry condition, eq. (\ref{Mnu_inv}), with $k=2$ and $S_\nu = \rho(S)$ given in eq. (\ref{G3_gen}).

It is evident from $M_\nu$ in eq. (\ref{Mnu_A4_w4}) and $M_l$ in eq. (\ref{Ml_A4}) that the ratio of solar to atmospheric squared mass difference and leptonic mixing parameters depend only on two complex parameters $g_2/g_1$ and $g_3/g_1$. Using the $\chi^2$ optimization technique and the results from the latest global fit to the neutrino oscillation parameters from NuFIT v5.3 (without SK atmospheric data) \cite{Esteban:2020cvm}, we investigate if they can lead to the viable neutrino masses and mixing. For the normal ordering in the neutrino masses, we find the best-fit value corresponding to $\chi_{\rm min}^{2} = 7.98 $ for 
\be \label{A4_sol_param}
g_2 / g_1 = 0.3398 + 0.2018\, i\,, \quad g_3/g_1 = -0.3251 - 0.0847\, i\,. \ee
In terms of the observable quantities, the solution leads to
\begin{gather*}
 r = \frac{\Delta m_{\rm sol}^{2}} {\Delta m_{\rm atm}^{2}} = 0.0296 \quad \sin ^{2} \theta_{12}=0.341, \quad \sin ^{2} \theta_{13}=0.022, \quad \sin ^{2} \theta_{23}= 0.578, \\
  \delta_{C P}= 1.239\, \pi , \quad \alpha_{21}= 1.102\, \pi, \quad \alpha_{31}= 0.983\, \pi,\\
m_{1}=  18.41\, \mathrm{meV} , \quad m_{2}= 20.32\, \mathrm{meV}, \quad m_{3}= 53.30\, \mathrm{meV}\, \\
m_{\beta \beta}= 18.34\, \mathrm{meV}\,.
\end{gather*}
The absolute values of the elements of the $U_{\rm PMNS}$ matrix are found as
 \begin{equation}
|U_{\rm{PMNS}}| =
\begin{pmatrix}
0.803 & 0.577 & 0.149 \\
0.318 & 0.577 & 0.752 \\
0.504 & 0.577 & 0.642
\end{pmatrix}\,,
\end{equation}
and the expected fixed column prediction from the residual symmetries can be seen. For the inverted ordering in the neutrino masses, we do not find acceptable $\chi^2_{\rm min}$ simultaneously being consistent with the upper bound on the effective neutrinoless double beta decay mass $m_{\beta \beta}$ \cite{KamLAND-Zen:2024eml} and on the sum of neutrino masses \cite{Planck:2018vyg,DESI:2024mwx}. The same is also observed in \cite{Novichkov:2018yse} earlier.

The generalization to higher even weights is straightforward. By utilizing the available irreps of modular Yukawa multiplets at a given weight, one can expand $M_\nu$ in the same manner as done in eq. (\ref{Mnu_A4_w4}). The number of invariant terms determines the number of free parameters. It is expected that a better fit to the data can be achieved if a sufficient number of these parameters are available. The fixed column prediction and resulting correlation shown in the left panel of Fig. \ref{fg1} remain valid, regardless of the number and values of these parameters.

It is also possible to reduce the number of parameters by considering specific ultraviolet realization of the Weinberg operator. For example, consider the underlying model extended by three singlet chiral superfields $\hat{N}^c$ transforming as $A_4$ triplet and with vanishing weight. The superpotential involving the interactions of these fields can be parametrized as
\be \label{WN}
W_N =  a\,Y^{(2)}_{\bf{3}}\,(\hat{L} \hat{N}^c)_{{\bf 3}_S} \hat{H}_u  + b\, Y^{(2)}_{\bf{3}}\,(\hat{L} \hat{N}^c)_{{\bf 3}_A} \hat{H}_u + M\, (\hat{N}^{c} \hat{N}^{c})_{\bf 1}\,.\ee
The above leads to the Dirac neutrino and right-handed Majorana neutrino mass matrices as:
\be \label{MD_MR}
M_D = v_u \left( \ba{ccc} 2 a Y_{1} & (-a + b) Y_{3}  & (-a-b) Y_{2}\\
 (-a - b) Y_{3} & 2 a Y_{2} &(-a + b) Y_{1}  \\
(-a + b) Y_{2} & (-a - b) Y_{1} & 2 a Y_{3}\\ \ea \right)\,,~M_R = M \left(\ba{ccc}
1 & 0  & 0\\
0 & 0 & 1 \\
0 & 1  & 0 \\ \ea \right)\,, \ee
respectively. The light neutrino masses induced by the type I seesaw mechanism is given by
\be \label{seesaw}
M_{\nu} = - M_{D}M_{R}^{-1} M_{D}^{T}\,. \ee
The neutrino mass matrix found using eqs. (\ref{seesaw},\ref{MD_MR}) can be shown as a specific case of the $M_\nu$ obtained in eq. (\ref{Mnu_A4_w4}) from the consideration of the Weinberg operator. Explicitly, identifying $\Lambda = M$ and
\beqa \label{gi_A4}
g_1 = \frac{2}{3}(a^{2} - 3 b^{2})\,,~~g_2 = -\frac{1}{3}(a^{2} + 6 a b - 3 b^{2})\,,~~g_3=-\frac{1}{3}  (a^{2} + 3 b^{2})\,,
\eeqa
we can relate the two.

The $M_\nu$ obtained in type I seesaw case is even more predictive as only complex parameter $b/a$ can determine $r$ and the leptonic mixing parameters.  In this case, we find $\chi_{\rm min}^{2} = 9.92$ for the normally ordered neutrinos, which is only marginally greater than the corresponding solution obtained for the most general Weinberg operator. The minimum of $\chi^2$ is obtained for 
\be \label{ab_A4}
{\rm Abs} [b / a] = - 3.652, \quad {\rm Arg}[b/a]= -0.074 \,.\ee
and it leads to
\begin{gather*}
r = \frac{\Delta m_{\rm sol}^{2}} {\Delta m_{\rm atm}^{2}} = 0.0295, \quad \sin ^{2} \theta_{12}=0.341, \quad \sin ^{2} \theta_{13}=0.022, \quad \sin ^{2} \theta_{23}=0.604, \\
 \delta_{C P}= 1.053\, \pi, \quad \alpha_{21}= 0.829\, \pi, \quad \alpha_{31}= 1.069\, \pi,\\
m_{1}=0, \quad m_{2}= 8.608\, \mathrm{meV}, \quad m_{3}=50.11\, \mathrm{meV}\,, \\
m_{\beta \beta}= 3.90\, \mathrm{meV}\,.
\end{gather*}
\hspace{0.0001cm}
Note that $M_D$ satisfies the anti-symmetry $S_\nu^T\,M_D\,S_\nu = - M_D$ as a consequence of eq. (\ref{bl_inv}) at $\tau=i$. This leads to ${\rm Det} (M_D)=0$ which ultimately gives rise to a massless neutrino as seen in the above solution. 

\begin{figure}[t!]
    \centering
    \includegraphics[width=0.65\textwidth]{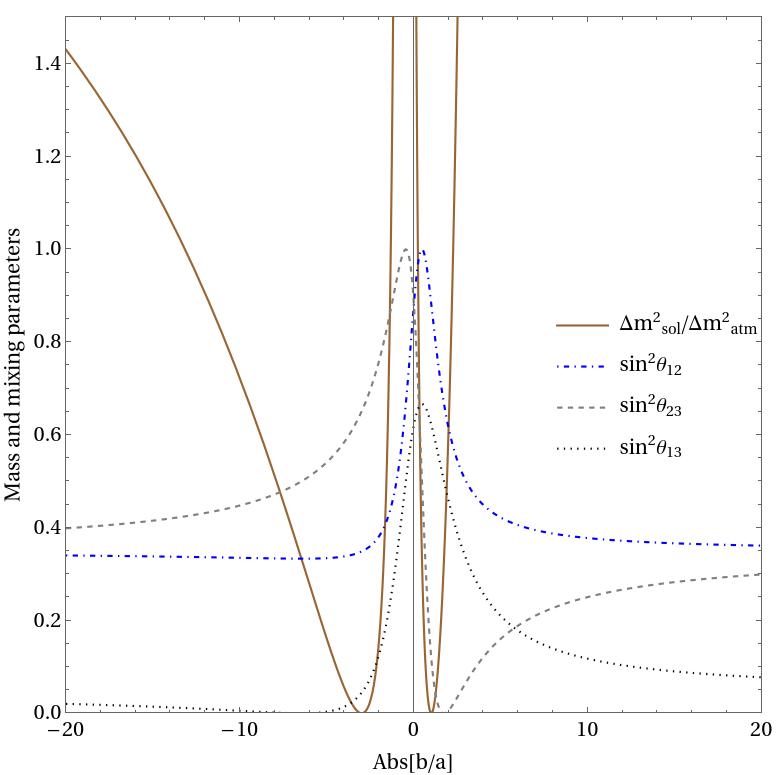}
    \caption{The variation of $\Delta m_{\rm sol}^{2}/\Delta m_{\rm atm}^{2}$ and the leptonic mixing angles  with respect to ${\rm Abs}[b/a]$ for a UV completion of $k_L=2$ Weinberg operator.}
    \label{fg5}
\end{figure}

It is remarkable that only two real parameters reproduce all the observables in a manner consistent with the data. The residual symmetry ensured by $\tau = i$ leads to a fixed column in the $U_{\rm PMNS}$ matrix and predicts a massless neutrino as specific, concrete, and verifiable outcomes of the setup. In Fig. \ref{fg5}, we also show the variation of $r$ and the mixing angles as a function of ${\rm Abs}[b/a]$. It can be observed that for a significantly large range of ${\rm Abs}[b/a]$ values, two of the mixing angles are large while one remains relatively small. Therefore, the observed pattern is a generic expectation from this setup. $r$ varies drastically and its small value is non-generic feature of this model. 
\subsection{Model based on $\Gamma_{4} \cong S_{4}$}
We now give an example of model based on $\Gamma_4$ with an aim to accommodate $|c_{\rm TL}|=\frac{1}{\sqrt{6}}(2,1,1)^T$ as the first column of the $|U_{\rm PMNS}|$.  The group $S_4$ possesses two inequivalent one-dimensional, a two-dimensional and two three-dimensional irreps. Our choices of representation and weight assignments for the lepton sector superfields and the flavon are shown in Table \ref{tab2}.
\begin{table}[h!]
    \centering
    \begin{tabular}{c c c c c}
    \hline
    \hline
   ~~~~~~~~~~~~ & ~~~$\hat{L}$~~~ & $\hat{E}^c =\{ \hat{e}^{c},\hat{\mu}^{c},\hat{\tau}^{c}\}$  & ~~~$\hat{H}_{u,d}$~~~ & ~~~$\hat{\Phi}$~~~ \\
     \hline
    $S_4$ & {\bf{3}} & {\bf{3}} & {\bf{1}} & {\bf{3}} \\
   weights & $k_L$ & $k_{l^c}$ & 0 & $-(k_L+k_{l^c})$  \\
    \hline
    \hline
    \end{tabular}
    \caption{Field content, choice of weights and charge assignments under $\Gamma_{4} \cong S_{4}$.}
    \label{tab2}
\end{table}

The superpotential involving the charged leptons can be written as
\be \label{}
W_l = \alpha\,(\hat{L} \hat{E}^c)_{\bf 1}\, \hat{H}_d + \frac{\beta^\prime}{\Lambda}\,(\hat{L} \hat{E}^c)_{{\bf 3}_S}\,\hat{\Phi}_{\bf 3}\, \hat{H}_d + \frac{\gamma^\prime}{\Lambda}\,(\hat{L} \hat{E}^c)_{{\bf 3}_A}\,\hat{\Phi}_{\bf 3}\, \hat{H}_d\,. \ee
The charged lepton mass matrix emerging from the above is required to possess a $Z_3$ symmetry corresponding to the generator 
\be \label{Tl_S4}
T_l = \left(
\begin{array}{ccc}
 0 & \frac{1}{\sqrt{2}} & \frac{1}{\sqrt{2}} \\
 \frac{i}{\sqrt{2}} & -\frac{i}{2} & \frac{i}{2} \\
 -\frac{i}{\sqrt{2}} & -\frac{i}{2} & \frac{i}{2} \\
\end{array}
\right)\,.\ee
The VEV configuration of flavon determined by $T_l\,\langle \Phi \rangle = \langle \Phi \rangle$, in this case, leads to
\be \label{flavon_vev_S4}
\langle \Phi \rangle = v_\Phi\, \left(1, \sqrt{i},-i\sqrt{i} \right)^{T}\,.\ee
The resulting charged lepton mass matrix is obtained as
\be \label{Ml_S4}
M_l = v_d\,\left( \ba{ccc}
\alpha & \sqrt{i}(\beta - i \gamma)  & i\sqrt{i}(\beta + i \gamma)\\
\sqrt{i}(\beta + i \gamma) & \beta & \alpha + \gamma \\
i\sqrt{i}(\beta - i \gamma) & \alpha - \gamma  & -\beta \\ \ea \right)\,, \ee
where $\beta = \beta^\prime v_\Phi/\Lambda$, $\gamma = \gamma^\prime v_\Phi/\Lambda$. The parameter $\alpha$ can be chosen real without loss of generality, while $\beta$ and $\gamma$ are complex. As we show later, the above $M_l$ can reproduce the desired charged lepton mass hierarchy due to the freedom offered by $\beta$ and $\gamma$. Also, the $M_l M_l^\dagger$ is diagonalized by a unitary matrix which does not depend on the values of $\alpha$, $\beta$ and $\gamma$ as dictated by residual symmetry condition, eq. (\ref{Ml_inv}).

\subsubsection{Weinberg operator with odd $k_L$}
For $k_L=1$, the neutrino mass matrix is made up of modular forms of weight $k_Y = 2 k_L =2$. The group $\Gamma_4$ possesses two modular multiplets with $k_Y=2$, namely $Y_{\bf 2}^{(2)}$ and $Y_{\bf 3}^{(2)}$, which transform as doublet and triplet of $S_4$, respectively. Using the general expansion, eq. (\ref{X_Y}), the neutrino mass matrix arising from the most general Weinberg operator with weight $2$ can be written as
\be \label{Mnu_S4_w2}
M_\nu = \frac{g_1 v_u^2}{\Lambda} \left( \ba{ccc}    2 (Y_{\textbf{2}}^{(2)})_{1} & 0 & 0 \\
    0 & \sqrt{3} (Y_{\textbf{2}}^{(2)})_{2} & - (Y_{\textbf{2}}^{(2)})_{1} \\
    0 & - (Y_{\textbf{2}}^{(2)})_{1} & \sqrt{3} (Y_{\textbf{2}}^{(2)})_{2} \\ \ea \right) \Bigg|_{\tau=i}\,. \ee
Even though $Y_{\bf 3}^{(2)}$ is present, it does not contribute into the $M_\nu$ due to vanishing CG coefficients. The values of $Y_{\bf 2}^{(2)}$ at $\tau=i$ can be obtained using the eigenvalue equation (\ref{Y_fp}). We find
\beqa \label{Yi_S4}
Y_{\bf 2}^{(2)}(i) &=& (Y_{\bf 2}^{(2)}(i))_1\,\left(1,-\frac{1}{\sqrt{3}}\right)^T\,.\eeqa
Substitution of these values in eq. (\ref{Mnu_S4_w2}) leads to ${\rm Det}(M_\nu)={\rm Tr}(M_\nu) =0$ as already anticipated. $M_\nu$ can also be made real without loss of generality and, therefore, it necessarily implies a massless and two degenerate neutrinos.

\subsubsection{Weinberg operator with even $k_L$}
The Weinberg operator of weight $k_Y = 2 k_L =4$ can be constructed in the similar ways. As it can be seen from \cite{Ding:2023htn}, there exists four modular multiplets: $Y_{\bf 1}^{(4)}$, $Y_{\bf 2}^{(4)}$, $Y_{\bf 3}^{(4)}$, $Y_{\bf 3^\prime}^{(4)}$. Only three of them can contribute to the neutrino masses due to their non-vanishing CG coefficients and we obtain:
\beqa \label{Mnu_S4_w4}
M_\nu &=& \frac{v_u^2}{\Lambda} g_1\Bigg[ \left( \ba{ccc}
    Y_{\textbf{1}}^{(4)} & 0 & 0 \\
    0 & 0 & Y_{\textbf{1}}^{(4)} \\
    0 & Y_{\textbf{1}}^{(4)} & 0 \\
    \ea \right) + \frac{g_2}{g_1} \left( \ba{ccc} 
     2 (Y_{\textbf{2}}^{(4)})_{1} & 0 & 0 \\
    0 & \sqrt{3} (Y_{\textbf{2}}^{(4)})_{2} & - (Y_{\textbf{2}}^{(4)})_{1} \\
    0 & - (Y_{\textbf{2}}^{(4)})_{1} & \sqrt{3} (Y_{\textbf{2}}^{(4)})_{2} \\
  \ea \right)  \nonumber \\
    &+& \frac{g_3}{g_1} \left( \ba{ccc}
    0 & (Y_{\bf 3'}^{(4)})_{2} & - (Y_{\bf 3'}^{(4)})_{3} \\
    (Y_{\bf 3'}^{(4)})_{2} & (Y_{\bf 3'}^{(4)})_{1} & 0 \\
    - (Y_{\bf 3'}^{(4)})_{3} & 0 & - (Y_{\bf 3'}^{(4)})_{1} \\ \ea \right) \Bigg] \Bigg|_{\tau=i}\,.\, \eeqa
The $Y_{\bf 3}^{(4)}$ does not contribute to $M_\nu$ because of the symmetric nature of the Weinberg operator with respect to $\hat{L}$. The $M_\nu$ obtained at weight $4$ in case of $S_4$ and $A_4$ are therefore parameterized by the same number of free parameters. The values of $Y_{\bf 1}^{(4)}$, $Y_{\bf 2}^{(4)}$, $Y_{\bf 3^\prime}^{(4)}$ at $\tau=i$ can be obtained from the expressions given in \cite{Ding:2023htn}.

By performing the numerical analysis, we find that the $M_l$ in eq. (\ref{Ml_S4}) and the above $M_\nu$ lead to viable lepton sector observables with normal ordering in the neutrino masses for
\begin{gather*}
\beta/\alpha = (-3.53 \times 10^{-4}) + 0.5003\, i, \quad \gamma/\alpha= 0.975 - 0.0018\, i \\
g_2/g_1 = 0.1227 + 0.1754\, i, \quad g_3/g_1 = -1.630 + 1.049\, i \,.
\end{gather*}
An excellent fit corresponding to $\chi_{\rm min}^{2} = 2.47 $ is obtained in this case and all the observables are fitted within the $3\sigma$ range of their experimentally extracted values. At the minimum of $\chi^2$, they  are:
\begin{gather*}
\frac{m_{e}}{m_{\mu}} = 0.0048, \quad \frac{m_{\mu}}{m_{\tau}} = 0.0591,\quad r = \frac{\Delta m_{\rm sol}^{2}} {\Delta m_{\rm atm}^{2}} = 0.0297, \\
\sin ^{2} \theta_{12}=0.318, \quad \sin ^{2} \theta_{13}=0.022, \quad \sin ^{2} \theta_{23}=0.601, \\
 \delta_{C P}= 0.351\, \pi \quad \alpha_{21}= 0.345\, \pi, \quad \alpha_{31}=  0.857\, \pi,\\
m_{1}= 31.02\, \mathrm{meV} , \quad m_{2}= 32.19\, \mathrm{meV}, \quad m_{3}= 58.82\, \mathrm{meV}\,, \\
m_{\beta \beta}= 28.15\, \mathrm{meV} \,.
\end{gather*}
The lepton mixing matrix is given by
\begin{equation} 
|U_{\rm PMNS}|=
\left(
\begin{matrix}
0.816 & 0.558 & 0.148 \\
0.408 & 0.496 & 0.766 \\
0.408 & 0.665 & 0.625
\end{matrix}
\right)\,.
\end{equation}
The first column of the $U_{\rm PMNS}$ matrix is fixed by the residual symmetries of the neutrino and charged lepton mass matrices, independent of the values of other parameters. The obtained solution leads to a sum of neutrino masses $\sum m_i = 0.12$ eV, which aligns with the current cosmological upper limit \cite{Planck:2018vyg,DESI:2024mwx}. It is possible to achieve an smaller $\sum m_i$ with only a marginal increase in $\chi_{\text{min}}^{2}$. For example, we find $\chi_{\text{min}}^{2} = 12.9$ corresponding to $\sum m_i = 0.10 $ eV.

\section{Summary and Discussions}
\label{sec:7}
Recent studies have indicated that many models of neutrino masses and lepton mixing using modular invariance tend to favour modulus values close to the self-dual point, $\tau = i$. Motivated by this observation, we explore the structure of the light neutrino mass terms precisely at $\tau = i$ in this article. Assuming that neutrino masses originate entirely from the Weinberg operator and are modular forms, the neutrino mass matrix $M_\nu$ at $\tau = i$ is shown to possess a residual $Z_2$ flavour symmetry or antisymmetry. The latter arises when the modular weight of $M_\nu$ is $2k$, with $k$ being an odd integer. This implies that at least one of the neutrinos is massless. Additionally, the other two states must be degenerate if $M_\nu$ is real. These findings are independent of the underlying finite modular group.

The nature of the residual $Z_2$ symmetry or antisymmetry transformations acting on the neutrino flavours depends on the finite modular group and is uniquely determined for a given group. If the charged lepton mass matrix also possesses some residual symmetry, this structure can predict a column in the leptonic mixing matrix. The mass pattern in the case of antisymmetry, as well as the fixed column in both symmetry and antisymmetry cases, are direct outcomes of the residual symmetries preserved at $\tau = i$, independent of model parameters.

It is interesting that the modular-invariant frameworks with the modulus fixed at $\tau = i$ lead to structures that can be analysed using techniques based on residual symmetries, similar to those used in standard flavour symmetry considerations. By using this connection, we have derived a comprehensive set of predictions for a fixed column in the leptonic mixing matrix. The results, particularly for small finite modular groups, suggest that the elements of the lepton mixing matrix are not very hierarchical, potentially explaining why $\tau \approx i$ is favoured by most modular models of leptons. It is important to note that some observations made in this study have already been realized in earlier works based on specific modular models, as mentioned in the preceding sections. Nevertheless, the fact that these observations arise from a more general structure and can extend beyond specific models is a noteworthy aspect of this study.

The results based on the residual symmetries obtained in this work are not exact and are subject to corrections arising from various sources, as happens in most modular constructions. In general, the Kähler potential is not well constrained by modular invariance, and it can induce large corrections to the effective mass matrices through renormalization factors. These corrections are indeterminate in bottom-up approaches and, if large, can spoil the residual symmetries entirely. Therefore, one is forced to assume a minimal Kähler potential, with the expectation that it can be arranged with the help of additional symmetries \cite{Nilles:2020kgo} or may follow from some top-down approach \cite{Cvetic:1991qm}. Supersymmetry-breaking effects can also affect the specific predictions. These effects are studied in \cite{Criado:2018thu}, where it is shown that the corrections due to supersymmetry breaking can be made small if the breaking scale and the messenger scale are reasonably far apart. Supersymmetry breaking also contributes to threshold corrections, the strength of which depends on the exact spectrum of superpartners and the value of the ratio of the VEVs of the two Higgs doublets. The latter also affects the stability of the predicted values of masses and mixing parameters under renormalization group evolution, and depending on the absolute neutrino mass scale, the effects can be sizable or negligible. All these effects can be studied conclusively, and their impact on the predictions can be meaningfully assessed only in the context of specific models. Nevertheless, the analysis presented in this work helps identify the specific models based on the desired predictions at leading order.

\section*{Acknowledgements}
We are grateful to Anjan S. Joshipura for valuable comments. This work is supported by the Department of Space (DOS), Government of India. MK would like to acknowledge Physical Research Laboratory, Ahmedabad for the Post-Doctoral Fellowship. KMP acknowledges partial support under the MATRICS project (MTR/2021/000049) from the Science \& Engineering Research Board (SERB), Department of Science and Technology (DST), Government of India.


\bibliography{ref}
\end{document}